\documentclass{PoS}

\usepackage{amsmath} 
\usepackage{amssymb} 
\usepackage{bm}
\usepackage{braket}

\newcommand {\E}[1]{\times 10^{#1}} 
\newcommand {\e}[1]{\mathrm{~#1}} 
\newcommand{\mc}[1]{\mathcal{#1}}

\newcommand{\mrm}[1]{\mathrm{#1}}

\renewcommand{\Im}[0]{\mrm{Im}}

\title{Scalar diquark in $t\bar t$ production and constraints on Yukawa sector of grand unified theories}

\ShortTitle{Scalar diquark in top-antitop production and constraints on Yukawa sector of GUTs}

\author{\speaker{Nejc Ko\v snik}
\\
Laboratoire de l'Acc\'el\'erateur Lin\'eaire,
Centre d'Orsay, Universit\'e de Paris-Sud XI,
B.P. 34, B\^atiment 200,
91898 Orsay cedex, France\\
       J. Stefan Institute, Jamova 39, P. O. Box 3000, 1001 Ljubljana,
       Slovenia\\
        E-mail: \email{kosnik@lal.in2p3.fr}}

\author{Ilja Dor\v sner\\
        Department of Physics, University of Sarajevo, Zmaja od Bosne 33-35, 71000 Sarajevo, Bosnia and Herzegovina\\
        E-mail: \email{ilja.dorsner@ijs.si}}

\author{Jure Drobnak\\
       J. Stefan Institute, Jamova 39, P. O. Box 3000, 1001 Ljubljana, Slovenia\\
        E-mail: \email{jure.drobnak@ijs.si}}

\author{Svjetlana Fajfer\\
  Department of Physics, University of Ljubljana, Jadranska 19, 1000 Ljubljana, Slovenia\\
       J. Stefan Institute, Jamova 39, P. O. Box 3000, 1001 Ljubljana,
       Slovenia\\
        E-mail: \email{svjetlana.fajfer@ijs.si}}

      \author{Jernej F. Kamenik\\
        J. Stefan Institute, Jamova 39, P. O. Box 3000, 1001
        Ljubljana,
        Slovenia\\
        Department of Physics, University of Ljubljana, Jadranska 19, 1000 Ljubljana, Slovenia\\
        E-mail: \email{jernej.kamenik@ijs.si}} \abstract{A colored
        weak singlet scalar state with hypercharge 4/3 is one of the
        possible candidates for the explanation of the unexpectedly
        large forward-backward asymmetry in $t \bar t$ production as
        measured by the CDF and D0 experiments. We investigate the
        role of this state in a plethora of flavor changing neutral
        current processes and precision observables of down-quarks and
        charged leptons. Our analysis includes tree- and loop-level
        mediated observables in the K and B systems, the charged
        lepton sector, as well as the $Z \to b \bar b$ width. We
        perform a fit of the relevant scalar couplings. This approach
        can explain the $(g-2)_\mu$ anomaly while tensions among the
        CP violating observables in the quark sector, most notably the
        nonstandard CP phase (and width difference) in the $B_s$
        system cannot be fully relaxed. The results are interpreted in
        a class of GUT models which allow for a light colored scalar
        with a mass below 1 TeV.  
}

\FullConference{The 2011 Europhysics Conference on High Energy Physics, EPS-HEP 2011,\\
		July 21-27, 2011\\
		Grenoble, Rh\^one-Alpes, France}

\begin{document}

\section{Introduction}
\label{sec:intro}
%

Recent CDF and D\O \, results on the forward-backward asymmetry (FBA)
in top quark pair production have attracted a lot of attention and a
number of proposals have been made in order to explain all the
relevant observables (for a review see
e.g.~\cite{Westhoff:2011tq}). Among these, a colored scalar, called
here $\Delta$, with SM charges $(\overline{\bm{3}}, \bm{1}, 4/3)$
couples to two up-quarks, hence it is also called a diquark. With large
coupling $g_{ut}$ of $\Delta$ to $u_R$ and $t_R$ one can accommodate
most of the present measurements of FBA and production cross section
of $t\bar t$~\cite{Dorsner:2009mq,Gresham:2011pa}. Only
flavor-changing couplings of $\Delta$ to up-quarks are allowed and we
study also the remaining two couplings in charm meson mixing
observables as well as in dijet and single top quark production
measurements at Tevatron.

On the other hand, rare processes involving down-type quarks and
charged leptons have played an important role in revealing possible
signs of new physics at low energies. A prominent example is the
anomalous magnetic moment of the muon, whose most precise experimental
measurement~\cite{Bennett:2004pv} deviates from theoretical
predictions within the SM~\cite{Jegerlehner:2007xe} by about three
standard deviations. The scalar $\Delta$ could play an important role
in those processes by an independent set of leptoquark couplings
$Y_{\ell d}$ to right-handed charged leptons and down-quarks.

\section{Processes with up-quarks}
Yukawa couplings of $\Delta$ to up-quarks are
\begin{equation}
 \frac{g_{ij}}{2} \epsilon_{abc}
  \bar{u}_{ia} P_L u^C_{jb} \Delta^c + \rm{h.c.}\,,
\end{equation}
where $P_{L,R} = (1\mp \gamma_5)$, $i,j$ are flavor indices, $a,b,c$
are color indices, and the totally antisymmetric tensor
$\epsilon_{abc}$ is defined with $\epsilon_{123}=1$. Matrix $g$ must
be antisymmetric due to antisymmetry in color
contraction. 

\begin{figure}[!h]
  \centering
  \begin{tabular}{ccc}
  \includegraphics[width=0.39\textwidth]{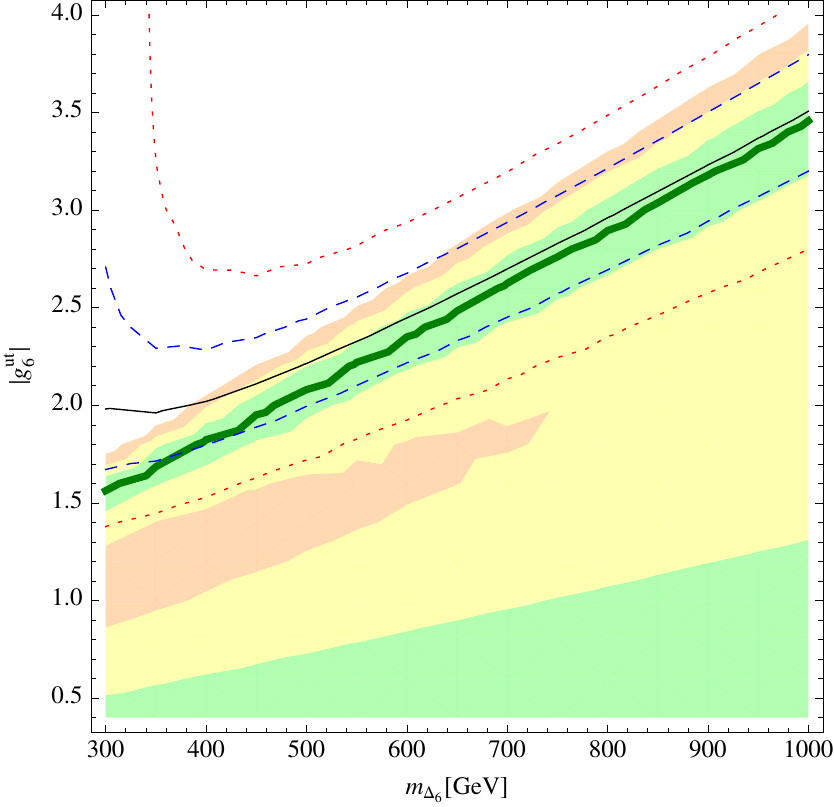}
  &&
  \includegraphics[height=.26\textheight]{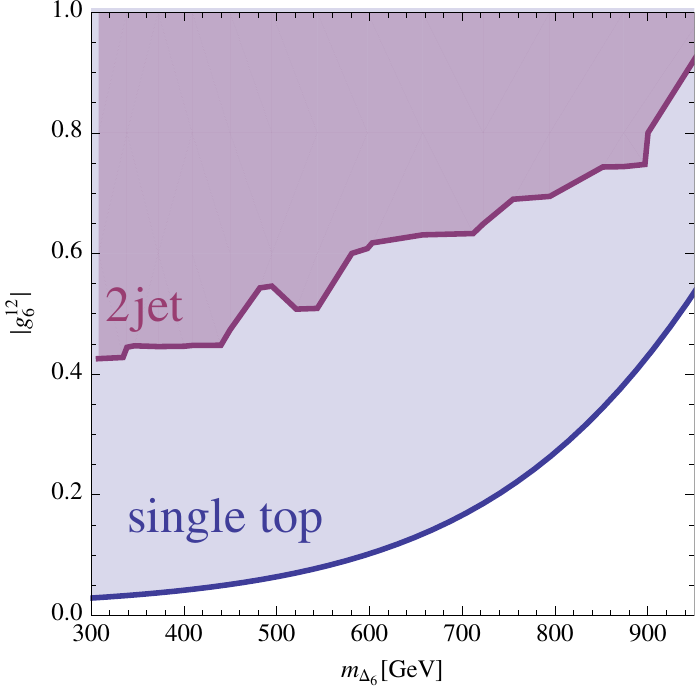}
  \end{tabular}
  \caption{Left: the 68, 95, 99\% CL regions in production
    cross section are shaded in green, yellow and orange
    respectively. The corresponding 68 (95)\% CL regions in the FBA
    are bounded by blue dashed (red dotted) contours. The best-fit
    contours are drawn in thick (thin) full lines for the cross
    section and the FBA respectively. Right: constraint on the $|g_{uc}|$ coupling
    and $\Delta$ mass from the single top production and di-jet at
    the Tevatron. The shaded areas are excluded.}
\label{fig:bounds}
\end{figure}

Exchange of $\Delta$ contributes to $t\bar t$ production amplitude in
the $u$-channel. It was emphasized in~\cite{Dorsner:2009mq} that
$\Delta$ at and below $1$\,TeV can enhance the SM prediction of the
forward-backward asymmetry $A_{FB}^{t\bar t}$ while not altering the
production cross section $\sigma_{t\bar t}$. This is achieved by
finding the parameter space of $m_{\Delta}$ and coupling $g_{ut}$,
contributing to partonic subprocess $u\bar u \to t\bar t$ in $p\bar
p$ collisions, which together with SM reproduce the measured values of
$A_{FB}^{t\bar t}$ and $\sigma_{t\bar t}$. The region where
experimental constraints can be satisfied within $1\sigma$ roughly
corresponds to a region, where mass of $\Delta$ and the coupling
$g_{ut}$ are correlated as~(Fig.~\ref{fig:bounds}, left graph)
\begin{equation}
  \label{eq:g13fit}
  |g_{ut}| = 0.9(2) + 2.5(4) \frac{m_{\Delta}}{1\e{TeV}}.
\end{equation}
Finite value of $g_{ut}$ allows us to constrain also the remaining two
couplings to up-quarks.

Potential contributions of $\Delta$ in $D-\bar D$ mixing can affect
the dispersive amplitude $M_{12}$ via box diagrams consisting of
$\Delta$ and $t$-quark.  Assuming $m_\Delta$ is between $300\e{GeV}$
and $1\e{TeV}$~\cite{Dorsner:2009mq} we can safely integrate out both
the top quark and $\Delta$ at a common scale $\mu = m_{\Delta}$ and
find an operator with right handed currents
\begin{equation}
  \label{eq:effham}
  \mc{H}^{\Delta C=2} =\frac{(g_{ut} g_{ct}^*)^2 h(m^2_{\Delta}/m^2_t)}{32 \pi^2 m_t^2} Q_6, \qquad Q_6 = (\bar u_R \gamma^\mu c_R) (\bar u_R \gamma_\mu c_R)\,.
\end{equation}
Using the HFAG averages~\cite{Barberio:2008fa} of $D-\bar D$ mixing
parameters and a set of formulas~\cite{Gedalia:2009kh,Grossman:2009mn}
that connect experimental parameters $x$, $y$, $|q/p|$, and $\phi$ to
the underlying theoretical parameters $M_{12}$, $\arg
(\Gamma_{12}/M_{12})$, we can constrain the magnitude of $M_{12}$.
The upper bounds at $95$\,\% ($2\sigma$) confidence level
are~\cite{Dorsner:2010cu}
\begin{align}
  \label{eq:x12s12bounds}
  2 |M_{12}|/\Gamma < 9.6\E{-3},\qquad 2 |\Im M_{12}|/\Gamma  < 4.4\E{-3}.
\end{align}
The imaginary part of $M_{12}$ and the resulting bound depends on a
relative phase between $g_{ct}$ and $g_{ut}$. A robust bound
\begin{equation}
  \label{eq:g23bound}
  |g_{ct}| < 0.0038,
\end{equation}
at $2\sigma$ CL comes from bound on $|M_{12}|$ and is independent of
complex phases in couplings. Here the central value for $|g_{ut}|$, as
given in Eq.~\eqref{eq:g13fit}, has been assumed.

The effects of coupling $g_{uc}$ could be observed in the CDF search
for resonances in the invariant mass spectrum of
dijets~\cite{0812.4036} as well as from the single top production
cross section measurements at the Tevatron~\cite{0908.2171}.  The
first measurement constrains the $|g_{uc}|$ coupling directly since
the process can be mediated by $\Delta$ through $s$-, $u$- and
$t$-channel exchange diagrams interfering with leading order QCD
contributions at the partonic level. As shown above, the $tc$ channel
is severely constrained by experimental results on $D-\bar D$
oscillations and can be neglected.  We compare our prediction for the
dijet spectra (for details see~\cite{Dorsner:2010cu}) with the
experimental results~\cite{0812.4036}.  We obtain the bounds on
$|g_{uc}|$ as a function of $m_{\Delta}$ by comparing the obtained
theoretical spectrum for given values of $\Delta$ parameters against
the experimental spectra. The results are shown as the purple shaded
area in the right graph of Fig.~\ref{fig:bounds}.

The single top production cross-section is sensitive to the product of
$|g_{uc} g_{ut}|$~\cite{Dorsner:2010cu}.  On the partonic level we
have a $u$-channel $u\bar u \to t \bar c$ and an $s$-channel $u c \to
t u$ contributions. A bound on $|g_{uc}|$ can then be obtained by
using the $t\bar t$ FBA preferred values of $|g_{ut}|$ in
Eq.~(\ref{eq:g13fit}).  We employ a conservative approach and only
compare NP contributions with the experimental error on the combined
Tevatron result for the total single-top production cross-section
(summed over both $t$ and $\bar t$) of $\sigma_{1t} =
2.76^{+0.58}_{-0.47}$~pb~\cite{0908.2171}. Excluded parameter is
shown as blue shaded area in the right graph of Fig.~\ref{fig:bounds}.

%
\section{Leptoquark processes}
\label{sec:pheno}
\label{sec:ewframe}
%
Yukawa couplings of $\Delta$ to down quarks and leptons are
\begin{equation}
\label{eq:lagr}
  Y_{ij} \bar{\ell}_{i} P_L
  d^C_{ja} \Delta^{a*} + \rm{h.c.}\,,
\end{equation}
Together with the diquark couplings, $g$, leptoquark couplings can
destabilize the proton. However, the tree-level dimension-6 proton
decay mediating operator is forbidden by the antisymmetry of $g$ in
flavor space.

The leptoquark couplings endow the scalar $\Delta$ with a potential to
cause large effects in (flavor changing) neutral current processes of
down quarks and charged leptons (see~\cite{Saha:2010vw} for a recent
analysis of scalar leptoquark constraints from $K$ and $B$
sectors). The couplings $Y_{ij}$ in Eq.~\eqref{eq:lagr} must therefore
pass constraints coming from many precisely measured or bounded from
above low energy observables.

\label{sec:treelevel}
Potentially most severe are the tree-level constraints on the $Y$
matrix. These are neutral $K$ meson decays ($K_L \to \mu^- \mu^+$,
$e^+ e^-$, $\mu^\pm e^\mp$; $K_S \to e^- e^+$, $\mu^+\mu^-$), neutral
$B_{(s)}$ meson decays ($B_{d(s)} \to \ell^- \ell'^{+}$,
$\ell^{(}{}'{}^{)} = e,\mu,\tau$), inclusive $B\to X_s \ell^+ \ell^-$,
exclusive semileptonic $B\to \pi \ell^+ \ell^{\prime-}$ and $B\to K
\ell^+ \ell^{\prime-}$, semileptonic tau decays ($\tau \to \mu
\eta,\mu K_S, e K_S,\mu \pi^0,e \pi^0$) and $\mu-e$ conversion on
nuclei. For more details about these constraints we refer reader
to~\cite{Dorsner:2011ai}.

Next we turn our attention to observables which are affected by
leptoquark couplings of $\Delta$ at the one-loop level. These are
$K-\bar K$ and $B-\bar B$ mixing amplitudes, LFV neutral current
processes like the radiative $\mu$ and $\tau$ decays ($\mu \to e
\gamma$, $\tau \to \mu \gamma$, $\tau \to e \gamma$), as well as
flavor diagonal observables, such as the anomalous magnetic moments of
leptons or the decay width of the $Z$ to $b\bar b$ pairs. The most
interesting observable in 1-loop category is the muon anomalous
magnetic moment $(g-2)_\mu$. In the recent years, the experimental
result on the anomalous magnetic moment of the muon $a_\mu \equiv
(g-2)_\mu/2$ from BNL~\cite{Bennett:2004pv} has been about $3\sigma$
above the SM predictions~\cite{Jegerlehner:2007xe}
\begin{subequations}
\begin{align}
 a_\mu^\mrm{exp} &= 1.16592080(63)\E{-3}\,,\\
 a_\mu^\mrm{SM} &= 1.16591793(68)\E{-3}\,.
\end{align}
\end{subequations}
Treating both experimental and theoretical uncertainties as Gaussian, we
may identify the missing contribution to $a_\mu$
\begin{equation}
\delta a_\mu = a_\mu^\mrm{exp} - a_\mu^\mrm{SM} = (2.87\pm 0.93)\E{-9}\,,
\end{equation}
with the presence of NP.
\begin{figure}[h]
  \centering
  \begin{tabular}{lcccr}
    \includegraphics[width=.27\textwidth]{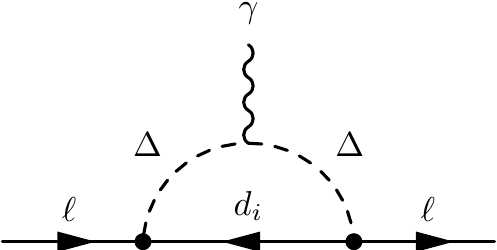} &&&&   \includegraphics[width=.27\textwidth]{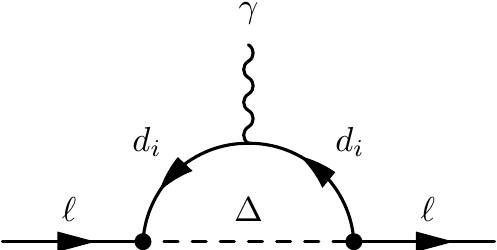} 
  \end{tabular}
  \caption{Diagrams with $\Delta$ and down-quarks contributing to
    the lepton anomalous magnetic moments.}
  \label{fig:aell}
\end{figure}
On top of the $a_\mu^\mrm{SM}$ we get a new contribution
\begin{equation}
a_\mu = \frac{1}{16\pi^2}\,\frac{m_\mu^2}{m_{\Delta}^2}\,\sum_{i=d,s,b} |Y_{\mu i}|^2\,,
\end{equation}
which comes from a diagram on Fig.~\ref{fig:aell}. Thus a finite magnitude is
preferred for a combination of the second row elements of $Y$
\begin{equation}
\label{eq:amu}
 \sum_{i=d,s,b} |Y_{\mu i}|^2 = (4.53 \pm 1.47)
 \E{-7}\times\frac{m_{\Delta}^2}{m_\mu^2} = (6.45\pm 2.09)\times \frac{m_\Delta^2}{(400\e{GeV})^2}\,.
\end{equation}

To see how the observed $a_\mu$ is generated while the FCNC and LFV
constraints are evaded we determine matrix $Y$ in a fit to all
abovementioned observables~\cite{Dorsner:2011ai} at a representative
mass $m_\Delta = 400\e{GeV}$.  In a trivial case, when we set $Y=0$ to
recover the SM, we find $\chi^2_\mrm{min} = 12.5=9.5_{a_\mu} +
1.5_\mrm{CKM} + 0.8_{\Delta m_s} + 0.3 + \cdots$ with a dominant
contribution from the $a_\mu$ anomaly. Finally we let $Y$ take any
value and we find a global minimum $\chi^2_\mrm{min} = 2.5 =
1.8_\mrm{CKM}+0.4_{\Delta m_s} + 0.14_{\sin 2\beta}\cdots$ for $15$
degrees of freedom, which signals a very good agreement of all
predictions with the considered observables. In particular, the best
point perfectly resolves the anomalous magnetic moment constraint
$a_\mu$ and slightly improves the quark flavor constraints. The
allowed $1\,\sigma$ ranges of $Y$ matrix elements are shown below
\label{eq:Yranges}
\begin{eqnarray}
\hspace{-1cm}|Y^\mrm{(1\,\sigma)}| &\in&
\begin{pmatrix}
<1.4\E{-6} & <8.7\E{-5} & < 4.2\E{-4}\\
<3.6\E{-3} \cup [2.1,2.9]  & <3.6\E{-3} \cup [2.1, 2.9] & <6.2\E{-4} \cup [2.3,2.7]\\
<5.6\E{-3} & <8.1\E{-3} & <9.6\E{-3}
\end{pmatrix}\,.
\end{eqnarray}
Couplings to the electron are strongly suppressed, while couplings to
the muon (the second row of $Y$) can take values of order $1$, in
order to satisfy the $a_\mu$ constraint. In the last row, elements
$Y_{\tau s}$ and $Y_{\tau b}$ can also be of order $0.01$ at
$1\,\sigma$.  Three possible regimes emerge in the second and the
third row, depending on which element in the second row is
large. These are
\begin{equation}
  \label{eq:Yhierarchy}
  \begin{pmatrix}
   &&\\
  \blacksquare & & \\
  \bullet & \bullet &\bullet
    \end{pmatrix}\,,\qquad
  \begin{pmatrix}
   &&\\
   &\blacksquare & \\
   \bullet& \bullet& \bullet
    \end{pmatrix}\,,
    \qquad
  \begin{pmatrix}
   &&\\
   & &\blacksquare  \\
   \bullet& \bullet&\bullet
    \end{pmatrix}\,.
\end{equation}
Here $\blacksquare$ stands for order $1$ element, $\bullet$ for (at
most) order $0.01$ element, and we neglect elements which are
$\lesssim 10^{-3}$.

%
\section{Conclusions}
%
Scalar $\Delta$ couplings to diquarks or to leptons and quarks exhibit
hierarchy, with $g_{ut}$ and $Y_{\mu i}$ sizable to explain the large
forward-backward asymmetry in $t\bar t$ production and anomalous
magnetic moment of the muon, respectively. Remaining couplings are
suppressed by $D-\bar D$ mixing ($g_{ct}$), dijet and single $t$
production ($g_{uc}$) or a combination of low-energy LFV and FCNC
bounds (all but one element $Y_{\mu i}$). For a leptoquark couplings
we also found that LFV $B$ and $\tau$ decay constraints lead to
strong limits on the tau lepton couplings to down quarks which in turn
exclude the possibility of sizable effects in the $B_s$
system~\cite{Dorsner:2011ai}.

The requirement of one large element $Y_{\mu i}$ puts $\Delta$ in the
so-called second generation leptoquark category. Moreover, as it does not couple
to neutrinos, the bound extracted from the recent LHC data for the
second-generation leptoquarks~\cite{Aad:2011uv} is truly applicable in
this case and reads $m_{\Delta} \gtrsim 380$\,GeV, accounting for the
reduced $\Delta \to \mu j$ branching ratio of order $\mathcal B\gtrsim
0.7$ due to the presence of the $\Delta \to t j$ decay
channel~\cite{Dorsner:2009mq}. This and the upper bound on its
mass---$m_{\Delta} < 560$\,GeV---that originates from simple
perturbativity arguments~\cite{Dorsner:2011ai} thus place it in a very
narrow window of discovery.

We have also systematically implemented~\cite{Dorsner:2011ai} all the
phenomenological constraints in a class of $SU(5)$ models where all
the fermion masses are generated at the tree-level to find out that
the explanation of the $a_\mu$ anomaly requires the vacuum expectation
value of the $45$-dimensional representation to be of the order of
$10^{-1}$\,GeV. This result implies that the up-quark couplings, in
this setup, are symmetric in nature.  We have
also shown that the symmetric scenario for the Yukawa couplings in the
down-quark and charged lepton case is not compatible with the
constraints due to the presence of light $\Delta$ and discussed
implications for the $SO(10)$ type of unification. The simplest of
possible realizations of both $SO(10)$ and $SU(5)$ with the symmetric
Yukawa sector, that could accommodate observed fermion masses, are
shown not to be viable unless $\Delta$ is heavy enough not to play any
role in low-energy phenomenology.

\end{document}